\documentclass[superscriptaddress,twocolumn,prb,
showpacs,floatfix,fleqn,longbibliography]{revtex4-1}
\usepackage[dvips]{graphicx}
\usepackage{epsfig} 
\usepackage{color}
\usepackage[normalem]{ulem}
\usepackage{amsmath,amsfonts,amssymb,bm}
\setcitestyle{numbers,square}
\UseRawInputEncoding

\begin{document}
\title{Superdiffusion in random two dimensional system with time-reversal symmetry and long-range  hopping}
\author{Xiaolong Deng} 
\affiliation{Leibniz-Rechenzentrum, Boltzmannstr. 1, D-85748 Garching bei M\"unchen, Germany}
\author{Ivan M.  Khaymovich} 
\affiliation{Stockholm University and KTH Royal Institute of Technology, 
Hannes Alfv\'ens v\"ag 12, SE-106 91 Stockholm, Sweden}
\author{Alexander L. Burin}
\email[]{aburin@tulane.edu}
\affiliation{Tulane University, New
Orleans, LA 70118, USA}
\date{\today}
\begin{abstract}
Although it is recognized that Anderson localization  takes place for all states at a dimension $d$ less  or equal $2$, while delocalization is expected  for hopping $V(r)$ decreasing with the distance slower or as $r^{-d}$, the localization problem in the crossover regime  for the dimension $d=2$ and hopping $V(r) \propto r^{-2}$ is not resolved yet. Following earlier suggestions we show that for the hopping determined by two-dimensional anisotropic dipole-dipole interactions in the presence of time-reversal symmetry there exist two distinguishable phases at weak and strong  disorder. The first phase is characterized by ergodic dynamics and superdiffusive transport, while the second phase is characterized by diffusive transport and delocalized eigenstates with fractal dimension less than $2$. 
The transition between phases is resolved analytically   using the extension of  scaling theory of localization  and verified numerically  using an exact numerical diagonalization. 
\end{abstract}

\maketitle


\section{Introduction.}
\label{sec:Intr}
 
Low dimensional systems with a dimension $d\leq 2$ possessing the time-reversal symmetry are critical in the  Anderson localization problem \cite{Anderson58}.  All states there must be exponentially localized at arbitrarily small disorder strength as was shown using the scaling theory of localization  \cite{Abrahams1979ScThLoc},  analysis of conductivity \cite{Gorkov792D} and extensive numerical simulations \cite{MacKinnon1981ScThLocNum,Eilmes1998Num2dLoc} (see also reviews \cite{LeeRamakrishnan1985RevModPhys,
ALTSHULERReview1985,EversMirlin08}). This localization is originated from the singular backscattering due to random potential dramatically enhanced in low dimension $d\leq 2$ where random paths inevitably return to the origin \cite{ALTSHULERReview1985}. 

The scaling theory of localization suggests the single parameter scaling for the dimensionless conductance $C=G/(e^2/\hbar)$ dependence on the size $L$ 
in the form  \cite{Abrahams1979ScThLoc,Gorkov792D}  ($G$ is the conductance) 
\begin{eqnarray}
\frac{d\ln(C)}{d\ln(L)}=\beta(C), 
\nonumber\\
\beta(C)= -\frac{c_{loc}}{C} + O(C^{-4}), ~ c_{loc}=\frac{1}{2\pi^2}, 
\label{eq:Band4}
\end{eqnarray}
where the $\beta$-function has been evaluated using expansion of the equivalent non-linear sigma-model  \cite{Wegner89RG,EversMirlin08} valid at $C\gg c_{loc}$. Eq. (\ref{eq:Band4})  predicts the reduction of conductance with the system size $L$ as $C(L)=C_{0}-c_{loc}\ln(L)$ where $C_{0}$ is the conductance at the lower cutoff  $L=1$.  Logarithmic reduction of conductance with the system size results in the   inevitable localization at large sizes $L$. 


This universal scaling  is limited to systems with a short-range hopping, while the hopping decreasing with the distance as $V(\mathbf{r})\propto r^{-d}$ or slower  leads to delocalization of all states  \cite{Anderson58,Yu1989PhononsFar,Levitov90,
Mirlin19961DLR,Cuevas_2007_maltfr,Fyodorov_2009}  except for the marginal case of diverging Fourier transform of a hopping amplitude  \cite{ab90RG,Modak_2016SPLR,Deng2018Dual,Cantin18PowLawIsotr,Kutlin20RGLoc,
ab21Scipost,Kutlin21EmFract,
Nosov2019AnomDelLR,Nosov2019CorrLongRange,
Deng19QuasiCr,
Roy21LongRangeQC,Tang2022nonergodic}.  If disorder is strong, eigenstates of the problem with the long-range hopping $V(\mathbf{r})\propto r^{-d}$ possess a multifractal  structure and the time dependent displacement of the particle $r$  obeys the law $r \propto t^{1/d}$, which is subdiffusive in $3D$ \cite{Levitov90}, diffusive in $2D$ and superdiffusive in $1D$ \cite{Mirlin19961DLR}.  
The long-range hopping $V(\mathbf{r})\propto r^{-2}$ is ubiquitous in  pure two-dimensional systems  \cite{YuLeggett88}, where it can be originated from the virtual exchange by two-dimensional photons leading to the $2D$ dipole-dipole interaction \cite{Aleiner112D} or indirect exchange by $2D$ electron-hole pairs leading to $2D$ RKKY interaction \cite{2DRKKFischer}.  Power-law distant dependent hopping  is crucial for the many-body localization problem  \cite{ab89MBLbreak,ab95TLSRelax,ab98book,ab98prl,Yao14MBLLongRange,ab15MBL,ab15MBLXY,
ab16GutmanMirlin,Nandkishore17lr,ab17SpectrDiffMBL,Singh17MBLLR,Halimeh17LR,
Homrighausen17MBLLR,Luitz2019MBLLR,Muller2019AlgebrLoc,DeTomasiMBLLR,
Halimeh2022MBLLR,Tikhonov2018MBLLR,Nag2019MBLLR,
ModakNag2020MBLLRCorrDis,
ModakNag2020RealSPRGMBL,DasSarma2022MBLLR,Monro16,Schultzen2022MBLLRDyn}, where long-range interaction can result in localization breakdown at arbitrary disorder.

For the hopping under consideration $r^{-d}$ and weak disorder,  there is the transition in $3D$ to the standard delocalized phase characterized by the diffusive transport and ergodic dynamics \cite{DengShlyap20163DDip}, while in $1D$ eigenstates turns out to be multifractals with the dimension smaller than $1$ \cite{Mirlin19961DLR}. $2D$ systems are more complicated,  because for the hopping  $V(\mathbf{r})\propto r^{-d}$ the dimensionless classical Drude conductance diverges logarithmically with the system size as $C_{0}(L)=c_{*}\ln(L)$ \cite{Aleiner112D}. Considering the balance of this logarithmic raise of conductance and its logarithmic suppression by coherent back scattering 
$c_{loc}\ln(L)$ Eq. (\ref{eq:Band4}), it was suggested in Ref. \cite{Aleiner112D} that two delocalized phases can exist including the superdiffusive ({\it fast}) phase at $c_{*}>c_{loc}$ and the {\it slow} phase with diffusive transport at $c_{*}<c_{loc}$ and the phase boundary realized at $c_{*}=c_{loc}$. Yet it turns out that in systems, possessing time-reversal symmetry, for isotropic dipoles considered in Ref. \cite{Aleiner112D} $c_{*}<c_{loc}$ for an arbitrarily disorder strength so the  {\it fast} phase does not exist. This is in contrast with the systems with a broken time-reversal symmetry, possessing the transition between fast and slow phases, characterized by the unstable fixed point \cite{Aleiner112D}. 

These achievements {\it motivated} us to search for the superdiffusive, {\it fast} phase using different hopping interaction including anisotropic dipole-dipole interaction with identically oriented dipoles.  Our preliminary estimates also show emergence of a superdiffusive phase for the $2D$ RKKY interaction at sufficiently weak disorder that needs a separate consideration.  These interactions  differ from the isotropic dipole  model of Ref. \cite{Aleiner112D} by the presence of dispersive modes with the mean free path increasing unlimitedly with decreasing disorder with similar increase of the logarithmic growth parameter $c_{*}$.  This  makes the appearance of the {\it fast}  phase with $c_{*}>c_{loc}$ unavoidable. similarly  the anisotropy-mediated localization investigated  in Ref. \cite{ab21Scipost} where the number of such modes is extensive,  though measure zero.  

The transition between phases occurring at  $c_{*}=c_{loc}$  differs qualitatively from the typical localization transition since in the present case the renormalization group equation for the conductance possesses the stable fixed point. Consequently, conductance approaches infinity with decreasing disorder strength in a continuous manner and the transition point can be expressed analytically through the single-particle green function. Consequently, we can find the transition point analytically for the model with the Lorentzian distribution of disorder where Green functions can be evaluated exactly, which is unprecendental for the localization problem. 
 
 This is in a sharp contrast with the standard localization transition in 3D systems with the short-range hopping \cite{AALRScalingLoc79,
LeeRamakrishnan1985RevModPhys} and the transition in the system similar to the present one but with a violated time-reversal symmetry \cite{Aleiner112D}. For those models conductance instantaneously jumps from a critical value to  infinity at the transition point, that can can be approached only numerically. 

Recent experimental realizations of $2D$ Anderson localization \cite{White2020Ultracold2D} and long-range hopping \cite{Lippe2021LRColdAtoms}   represent the steps  towards generating the settings targeted in the present work. Consequently, we believe that  its experimental realization is possible and it is strongly encouraged.  Two dimensional $r^{-2}$ dipole-dipole interaction emerges in high dielectric constant films in a limited distance range where superdiffusive behavior can be observed as discussed in the end of  Sec. \ref{sec:Sc}.

In addition to  the dipole-dipole interaction, there exists a long-range elastic $r^{-2}$ interaction in isolated dielectric films  with a similar angular dependence to that for a dipole-dipole interaction.  Similar phase transition is expected  for that interaction without  constraints like for a dipole-dipole interaction because elastic field  is located fully inside the material.  This is also true for an indirect exchange RKKY interaction within  two dimensional metals. Of course for all   interactions the consideration is limited to a subwavelength transport \cite{ab95TLSRelax}.


We investigate two phases for two-dimensional Anderson model, described in Sec.  \ref{sec:Model} with a long-range hopping, formulated below, using the extension of scaling theory of localization for the long-range hopping developed in Sec.  \ref{sec:ZerCond},  \ref{sec:Sc} and exact numerical diagonalization in Sec. \ref{sec:Num}.  The Lloyd model with the Lorentzian distribution of random site potentials is used since the Green functions can be evaluated exactly in this model \cite{Lloyd_1969}, and using them we can find analytically the transition between fast and slow phases. 

The phase boundary $c_{*}=c_{loc}$ and a finite-size scaling of conductance are identified analytically  in Secs. \ref{sec:ZerCond}, \ref{sec:Sc}.   The numerical study of Sec.  \ref{sec:Num} is targeted to check a consistency of the analytical theory with the results obtained by means of exact numerical diagonalization.   Our investigation of level statistics  in Sec.  \ref{subsec:LStat} shows the consistency of the analytically predicted phase boundary with the behavior of the level statistics   changing from the Wigner-Dyson statistics in the fast phase to the intermediate one  between Poisson and Wigner-Dyson otherwise.  The finite-size scaling of an eigenstate fractal dimensions reported in Sec. \ref{subsec:FractDim} is consistent with the analytical theory  under the assumption of a linear dependence of a fractal dimension on the inverse conductance, though the proportionality coefficient differs from the earlier theoretical predictions.  Finally, we demonstrate that a time-dependent displacement of a particle initially localized in a single site shows super diffusive behavior in accord with the theoretical expectations, although the direct comparison of analytical and numerical results is problematic because of an insufficient  maximum size of the system and the contribution of all eigenstates to the transport including the ``slow'' ones in the spectrum tails.  All numerical results would not be fully conclusive for the infinite size limit without the analytical theory, which is the main outcome of the present work. 


\section{Model}
\label{sec:Model}

Anderson model in $2D$ is investigated. The Hamiltonian of the model has the form 
\begin{eqnarray}
\widehat{H}=\frac{1}{2}\sum_{i,j}V_{ij}c_{i}^{\dagger}c_{j}+\sum_{i}\phi_{i}c_{i}^{\dagger}c_{i}, 
\label{eq:HamGen}
\end{eqnarray}
where the summation is performed over $N=L^2$ lattice sites enumerated by indices $i$ with coordinates $\mathbf{r}_{i}=(x_{i}, y_{i})$ occupying the periodic square lattice with a  period equal to  unity placed onto the surface of torus characterized by the radii $R=L/(2\pi)$.  Independent random energies $\phi_{i}$ obey the Lorentzian distribution 
\begin{eqnarray}
P(\phi) = \frac{1}{\pi}\frac{W}{W^2+\phi^2},  
\label{eq:RandPotLor}
\end{eqnarray}
having the width $W$ characterizing the disorder strength, while hopping amplitudes are given by the exchange of the dipolar excitations between interacting dipoles, oriented along the $x$-axis, via the interaction  \cite{Aleiner112D} 
\begin{eqnarray}
V(\mathbf{r}_{ij}) = V_0 \frac{x_{ij}^2 - y_{ij}^2}{r_{ij}^4}, \quad  r_{ij} = (x_{ij},y_{ij}). 
\label{eq:dipdip}
\end{eqnarray}
 These hopping terms are formed similarly to the interaction $(3x^2-r^2)/r^5$ in three dimensions \cite{LLFieldTh}.

The present model is different from that of Ref. \cite{Aleiner112D} because all transition  dipole moments are oriented along the  $x$ axis. To reproduce the settings of Ref.  \cite{Aleiner112D} we need to consider two degenerate states with identical random potentials in each site with perpendicular transition dipole moments and introduce the dipolar hopping between them accordingly.  Also a short range hopping can be added between states with identical transition dipole moments.  As noticed in the introduction in the present model there is the transition between fast and slow phases while it lacks in the isotropic model of Ref. \cite{Aleiner112D}. 

For the Lorentzian distribution of random potentials the Green's functions, averaged over the random potential realizations, can be evaluated exactly \cite{Lloyd_1969} 
in the momentum representation because averaging of local Green functions $1/(E-\phi_{i}-i\delta)$ with the distribution   Eq. (\ref{eq:RandPotLor}) yields $1/(E-iW)$ so averaging replaces all random potentials with the imaginary constant $iW$.  Then the  configurationally averaged Green function takes the form 
\begin{eqnarray}
G(E, \mathbf{q})=\frac{1}{E-V(\mathbf{q})-iW},
\nonumber\\ 
V(\mathbf{q})=\sum_{k}V_{jk}e^{i\mathbf{qr}_{jk}},
\label{eq:GrFun}
\end{eqnarray}
where $V(\mathbf{q})$ expresses the Fourier transform of the hopping $V_{ik}$.  For small wavevectors $q\ll 1$ it  can be approximated by the continuous limit of the dipole-dipole interaction Fourier transform with subtracted self-interaction as 
\begin{eqnarray}
V(\mathbf{q}) \approx 2\pi V_{0}\frac{q_{y}^2-q_{x}^2}{q^2}. 
\label{eq:FourTr}
\end{eqnarray}
as verified in Appendix \ref{sec:Modell} numerically, see also Ref. 
 \cite{ab21Scipost}. The Green functions are   needed for the calculation of a classical conductance given below.  There  we set $V_{0}=1$ and use the effective disorder strength parameter $W$ to distinguish different phases. 

\section{Classical conductance}
\label{sec:ZerCond}

 Long-range hopping results in a logarithmic divergence of a classical conductance. To characterize this divergence, let's  consider the generalized definition of a wavevector-dependent  conductance  \cite{Gorkov792D,
LeeRamakrishnan1985RevModPhys,Aleiner112D} that is a target of the renormalization group analysis. It reads 
\begin{widetext}
\begin{eqnarray}
C_{ab}(\mathbf{q}) = \int d\mathbf{p}  \int d\mathbf{p}_{1}   \frac{\partial V(\mathbf{p}+\mathbf{k}/2)}{\partial p_{a}} \left\langle {\rm Im} g(\mathbf{p}+\mathbf{k}/2, \mathbf{p}_{1}+\mathbf{k}/2){\rm Im} g(\mathbf{p}_{1}-\mathbf{k}/2, \mathbf{p}-\mathbf{k}/2)\right\rangle \frac{\partial V(\mathbf{p}_{1}-\mathbf{k}/2)}{\partial p_{1b}},
\label{eq:FullConduct}
\end{eqnarray}
\end{widetext} 
where $g(\mathbf{p}, \mathbf{p}_{1})$ stands for the  Green functions  taken at the  energy of interest $E$ before the configurational averaging $\langle...\rangle$.  
\begin{widetext}
\begin{center}
\begin{figure}[h]
\centering
\includegraphics[scale=0.6]{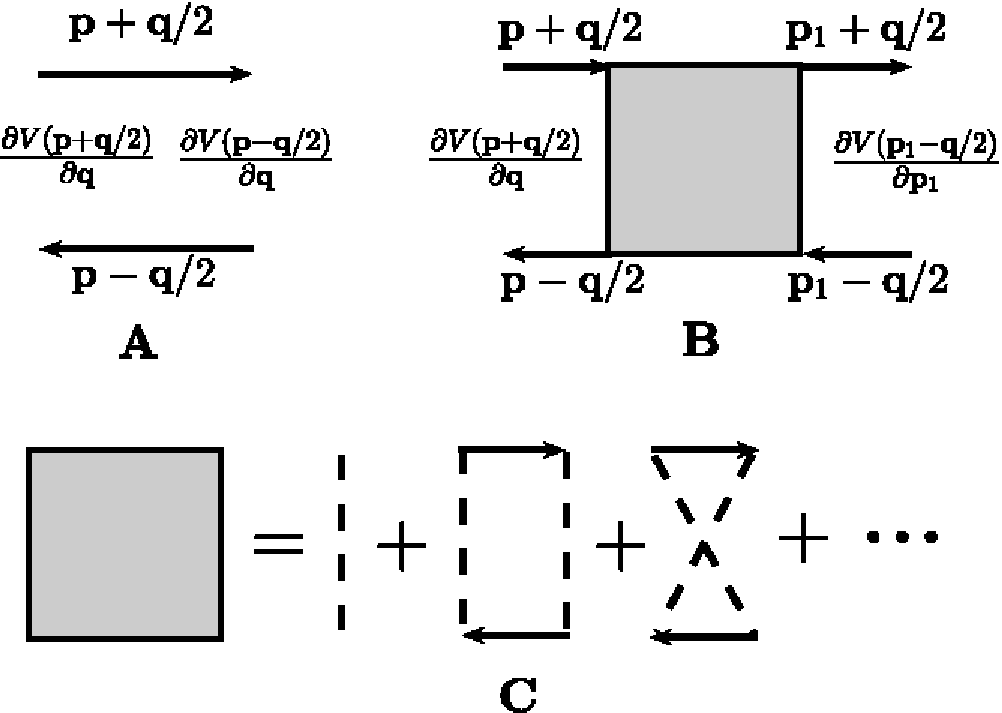}
\caption{\small   Classical (A) and quantum (B) and (C)  contributions to the dimensionless conductance. } 
\label{fig:DivContr}
\end{figure} 
\end{center}
\end{widetext}

The  conductance can be expressed using diagrams \cite{Gorkov792D,LeeRamakrishnan1985RevModPhys,
Altshuller1986Dim}  shown in  Fig. \ref{fig:DivContr}, where solid lines stand for configurationally averaged Green functions Eq.  (\ref{eq:GrFun}) and  dashed lines indicate correlations of two Green functions appearing due to the contributions of identical sites to both Green function.    This series represents the conductance expansion in the ratio of  the wavelength and the  mean free path thus giving  the quantum corrections  Fig. \ref{fig:DivContr}.B  to the classical conductance shown in Fig. \ref{fig:DivContr}.A.  For the classical conductance  input and output wavevectors are identical $\mathbf{p}=\mathbf{p}_{1}$.  This classical conductance  has a logarithmic divergence  due to the squared group velocity terms like  $\frac{\partial V(\mathbf{p})}{\partial p_{a}}^2 \propto p^{-2}$, Eq. (\ref{eq:FourTr})  lacking for the short-range hopping.  


For the quantum corrections to the conductance shown  in Fig. \ref{fig:DivContr}.B $\mathbf{p}\neq \mathbf{p}_{1}$ and, therefore,  no logarithmic divergence emerges.  Consequently,  the correlations between two Green functions are neglected, when defining a positive logarithmically diverging contribution to the   conductance, which we are  interested in.   This logarithmically diverging  part is evaluated below exactly since in the present model exact Green functions are known Eq. (\ref{eq:GrFun}).  

The classical dimensionless conductance tensor at given energy $E$ takes the form (see Eq.  (\ref{eq:FullConduct}) and Refs.   \cite {Gorkov792D,ALTSHULERReview1985,Aleiner112D})
\begin{widetext}
\begin{eqnarray}
C^{(0)}_{ab}(\mathbf{q})= \int \frac{d\mathbf{p}}{\pi (2\pi)^{d}} \left(\frac{\partial V(\mathbf{p}+\mathbf{q}/2)}{\partial p_{a}}\right)
\left(\frac{\partial V(\mathbf{p}-\mathbf{q}/2)}{\partial p_{b}}\right){\rm Im} G(E, \mathbf{p}+\mathbf{q}/2){\rm Im} G(E, \mathbf{p}-\mathbf{q}/2), 
\label{eq:SummCondS}
\end{eqnarray}
\end{widetext}
where $G(E, \mathbf{q})$ is  the retarded Green function at energy $E$ and wavevector $\mathbf{q}$  defined in Eq. (\ref{eq:GrFun}). 

The conductance in Eq. (\ref{eq:SummCondS}) diverges logarithmically at small wavevectors $p$ corresponding to  long distances. This divergence is caused by  the divergence of the mean squared displacement $\int d^2\mathbf{r} V(r)^2r^2$ within the Fermi-Golden rule approach. At long distances the Fourier transform continuous representation becomes exact so the use of the hopping amplitude Fourier transform in the form of Eq. (\ref{eq:FourTr}) is completely justified. 
Below we evaluate analytically the diverging part needed for the characterization of the phase transition for the anisotropic dipole - dipole  hopping.

For the dipole-dipole hopping Eqs. (\ref{eq:dipdip}), (\ref{eq:FourTr}) the integral for the classical conductance Eq. (\ref{eq:SummCondS}) diverges logarithmically at $p=0$. Since this divergence takes place within the domain $q<p<1$
(for meaningful wavevectors $q>1/L$) the conductance tensor diverging components in  Eq. (\ref{eq:SummCondS}) can be evaluated with the logarithmic accuracy as $C_{xx}=c_{x}\ln(1/q)$, $C_{yy}=c_{y}\ln(1/q)$, $c_{xy}=c_{yx}=0$, where 
\begin{widetext}
\begin{eqnarray}
c_{x}=\frac{\partial C^{(0)}_{xx}}{\partial \ln(L)}= \frac{4}{\pi}\int_{0}^{2\pi}d\phi \frac{\cos(\phi)^2\sin(\phi)^4W^2}{\left[(E-\pi \cos(2\phi))^2+W^2\right]^2}
\nonumber\\
=
\frac{(\pi^2-W^2-E^2){\rm Im}(\sqrt{E^2-(\pi-iW)^2})+\pi W{\rm Re}(\sqrt{E^2-(\pi -iW)^2})}{4\pi^3 \sqrt{(E+\pi )^2+W^2}},
\nonumber\\
c_{y}=\frac{\partial C^{(0)}_{yy}}{\partial \ln(L)}=\frac{4^2}{\pi}\int_{0}^{2\pi}d\phi \frac{\cos(\phi)^4\sin(\phi)^2 W^2}{\left[(E-\pi \cos(2\phi))^2+W^2\right]^2}
\nonumber\\
=\frac{(\pi^2 -W^2-E^2){\rm Im}(\sqrt{E^2-(\pi -iW)^2})+\pi W{\rm Re}(\sqrt{E^2-(\pi -iW)^2})}{4\pi^3 \sqrt{(-E+\pi )^2+W^2}}.
\label{eq:DrudeDipDip1}
\end{eqnarray}
\end{widetext}
In the middle of the band $E=0$ the conductance is isotropic, $C_{xx}=C_{yy}=c_{*}\ln(L)$, and  the logarithmic growth rate  $c_{*}$ is given by
 \begin{eqnarray}
c_{*}=\frac{1}{2W\sqrt{\pi^2 +W^2}}.
\label{eq:Sigm0dipE0r}
\end{eqnarray}

It is noticeable, that in the weak disorder limit ($W\rightarrow 0$) the rate parameter  $c_{*}$ in Eq. (\ref{eq:Sigm0dipE0r}) approaches infinity, so the transition to the superdiffusive regime should take place at a finite disorder strength $W$ where $c_{*}(W)=c_{loc}=1/(2\pi^2)$ in contrast  with the  isotropic dipole-dipole hopping \cite{Aleiner112D}.

The generalization to the arbitrary  distribution of random potentials can be made by the  replacements $E \rightarrow E-{\rm Re}\Sigma(E, 0)$ and $W \rightarrow -{\rm Im}\Sigma(E, 0)$ in Eqs. (\ref{eq:DrudeDipDip1}), (\ref{eq:Sigm0dipE0r}), where $\Sigma(E, 0)$ is the self-energy evaluated at energy $E$ and wavevector $q=0$. One should notice that $\Sigma(0, 0)=0$. Finding the self-energy  for arbitrary  distribution of random potentials remains a challenge; yet this problem is much easier compared to the localization problem itself. The classical  conductance serves as an input  to the renormalization group equation for the conductance derived below in Sec. \ref{sec:Sc}.  

\section{Renormalization group equation. } 
\label{sec:Sc}

Here we derive the $\beta$-function determining the size dependence of conductance in Eq. (\ref{eq:Band4}) within the one-loop order. The derivation below is given for the isotropic regime of symmetric conductances $c_{xx}=c_{yy}$ while for the anisotropic regime we give the results   in the end of the present section.  The isotropic regime is approximately valid for the  system under consideration at zero energy (see  Eq. (\ref{eq:Sigm0dipE0r}) in   Sec. \ref{sec:ZerCond}).  

We examine  the renormalization of conductance $C(q, p_{1})$ for   the orthogonal (possessing the time-reversal symmetry) sigma model  within the one-loop order assuming that the conductance is much greater than one, which is true near the transition point, where it approaches infinity.  Here $q$ is the current momentum and $p_{1}$ is  the maximum momentum  \cite{Wegner89RG} reduced during renormalization procedure.  The renormalization of the conductance  is associated with the reduction of the maximum momentum to the new value $p_{2} \ll p_{1}$.  
This renormalization  can be expressed as \cite{Wegner89RG,Mirlin19961DLR} ($q\ll p$)
\begin{widetext}
\begin{eqnarray}
q^2(C(q, p_{2})-C(q, p_{1})) \approx -\int'\frac{d\mathbf{k}}{\pi (2\pi)^2}\frac{C(|\mathbf{k}+\mathbf{q}|, p_{1})(\mathbf{k+q})^2-C(k, p_{1})k^2}{C(k, p_{1})k^2},  
\label{eq:CondIsCorr1Loop}
\end{eqnarray}
\end{widetext}
where integration $\int'$ is taken over the domain of momenta $p_{2}<p<p_{1}$ that is getting excluded  during the renormalization procedure. This is the one-loop order correction to the conductance similar to Eq.  (25) in Ref.  \cite{Mirlin19961DLR}, where it was considered for a one dimensional model with the long-range hopping. The terms $C(k, p_{1})k^2$, $C(|\mathbf{k}+\mathbf{q}|, p_{1})(\mathbf{k+q})^2$ in the denominators  are identical to the terms $|q+k|^{\sigma}$ and $|k|^{\sigma}$ in Ref. \cite{Mirlin19961DLR}. 


The initial conditions to  Eq.  (\ref{eq:CondIsCorr1Loop}) at large $p_{1} = O(1)$ are set using the classical order conductance as 
\begin{eqnarray}
C(q, 1) = c_{*}\ln(1/q), 
\label{eq:InCond}
\end{eqnarray}
where the inverse wavevector $q$ serves as the cutoff radius in the definition of the conductance. The long-range interaction enters into the consideration through this initial condition. 

In the limit $q \ll p_{1}$ one can expand the expression in the numerator to the second order in $q$ as (the first order disappears because of the integration over angles) 
\begin{widetext}
\begin{eqnarray}
C(q, p_{2})-C(q, p_{1}) \approx -\sum_{\alpha,\beta=x,y}\frac{q_{\alpha}q_{\beta}}{2q^2}\int'\frac{d\mathbf{p}}{\pi (2\pi)^2}\frac{\partial^2 (C(p, p_{1})p^2)}{\partial p_{\alpha}\partial p_{\beta}}\frac{1}{C(p, p_{1})p^2}. 
\label{eq:CondIsCorr1Loop1}
\end{eqnarray}
Evaluating derivatives and  averaging over angles of vector $\mathbf{p}$  we get
 \begin{eqnarray}
C(q, p_{2})-C(q, p_{1}) = -\int'\frac{d\mathbf{p}}{\pi (2\pi)^2}\left[C(p, p_{1})+\frac{\partial C(p, p_{1})}{\partial \ln(p)}+\frac{1}{4}\frac{\partial^2 C(p, p_{1})}{\partial \ln(p)^2}\right]
\frac{1}{C(p, p_{1})p^2}. 
\label{eq:CondIsCorr1Loop2}
\end{eqnarray}
\end{widetext}
Assuming that the logarithmic derivatives of the conductance are smooth functions (this is justified by the logarithmic size dependence of conductance in the initial condition Eq. (\ref{eq:InCond}) and  can be verified using the solution of the equation) one can perform logarithmic integration in the right hand side of Eq. (\ref{eq:CondIsCorr1Loop2}) and express this equation in the standard differential form similarly to Ref.  \cite{Wegner89RG}
\begin{widetext}
 \begin{eqnarray}
 \frac{\partial C(q, p)}{\partial \ln(p)}=\frac{1}{2\pi^2}\left[1+\frac{1}{C(p, p)}\frac{\partial C(p_{1}, p)}{\partial \ln(p_{1})}\bigg\rvert_{p_{1}=p}+\frac{1}{4C(p, p)}\frac{\partial^2 C(p_{1}, p)}{\partial \ln(p_{1})^2}\bigg\rvert_{p_{1}=p}\right]. 
\label{eq:RG1app}
\end{eqnarray}
\end{widetext}
Since the right hand side of Eq. (\ref{eq:RG1app}) is independent of the wavevector $q$ one can evaluate logarithmic derivatives using the initial condition Eq. (\ref{eq:InCond}) as 
$$
\frac{\partial C(p_{1}, p)}{\partial \ln(p_{1})}\bigg\rvert_{p_{1}=p}=-c_{*}, ~ \frac{\partial^2 C(p_{1}, p)}{\partial^2 \ln(p_{1})}\bigg\rvert_{p_{1}=p}=0. 
$$
Then Eq. (\ref{eq:RG1app}) takes the form 
 \begin{eqnarray}
 \frac{\partial C(q, p)}{\partial \ln(p)}=\frac{1}{2\pi^2}\left[1-\frac{c_{*}}{C(p, p)}\right]. 
\label{eq:RG2app}
\end{eqnarray}

The renormalized conductance at the given momentum $p$ can be determined  with the logarithmic accuracy as $C(p, p)$  and it can be denoted as $C(p)$ for the convenience. Using the initial condition Eq. (\ref{eq:InCond})  for the derivative with respect to the first argument we end up with the renormalization group equation in the form 
 \begin{eqnarray}
 \frac{d C(p)}{d \ln(p)}=-c_{*}+\frac{1}{2\pi^2}\left[1-\frac{c_{*}}{C(p)}\right]. 
\label{eq:RG2FinApp}
\end{eqnarray}
For the size $L$ dependent conductance one can express the relevant wavevector $p$ as $\eta_{1}/L$ for $\eta_{1}= O(1)$. This leads to the renormalization group equation for the size dependent conductance in the form 
 \begin{eqnarray}
 \frac{d C}{d \ln(L)}=c_{*}-c_{loc} +\frac{c_{*}c_{loc}}{C} +O(C^{-2}), 
\nonumber\\ 
 c_{loc}=\frac{1}{2\pi^2}.
\label{eq:RG2FinAppL2}
\end{eqnarray}
 Assuming that $C\gg c_{loc}$ we can ignore higher order terms. Then for $c_{loc}>c_{*}$ the steady state solution reads 
 \begin{eqnarray}
C=\frac{c_{*}c_{loc}}{c_{loc}-c_{*}} .  
\label{eq:RG2FinAppL}
\end{eqnarray}
It is a stable fixed point. This solution is applicable in the infinite-size limit and for $c_{loc}-c_{*} \ll c_{loc}$ where higher order terms in $1/C$ can be neglected.  
In the opposite case $c_{*}>c_{loc}$ the solution approaches infinity for $L\rightarrow \infty$. 

Formally Eq. (\ref{eq:RG2FinAppL2}) goes beyond the one-loop order expansion since the term inversely proportional to the conductance is comparable to the two-loop order contributions. However, since there is no contributions to the conductance up to four-loop order \cite{Wegner89RG}, we believe that we do not need to go beyond the one-loop order. 


The renormalization group equation for the anisotropic conductance can be derived similarly to the isotropic regime. In the one-loop order we got 
\begin{widetext}
\begin{eqnarray}
\frac{dC_{x}}{d\ln(L)}=c_{x}^{*}-c_{loc}\frac{C_{x}}{\sqrt{C_{x}C_{y}}}+\frac{2c_{x}^{*}}{\sqrt{C_{x}}(\sqrt{C_{x}}+\sqrt{C_{y}})} +\frac{C_{y}c_{x}^{*}-C_{x}c_{y}^{*}}{4\sqrt{C_{x}C_{y}}(\sqrt{C_{x}}+\sqrt{C_{y}})^2}+O(C^{-2}),
\nonumber\\
\frac{dC_{y}}{d\ln(L)}=c_{y}^{*}-c_{loc}\frac{C_{y}}{\sqrt{C_{x}C_{y}}}+\frac{2c_{y}^{*}}{\sqrt{C_{y}}(\sqrt{C_{x}}+\sqrt{C_{y}})} +\frac{C_{x}c_{y}^{*}-C_{y}c_{x}^{*}}{4\sqrt{C_{x}C_{y}}(\sqrt{C_{x}}+\sqrt{C_{y}})^2}+O(C^{-2}).
\label{eq:CondCorr1LoopAnis}
\end{eqnarray}
\end{widetext}

Similarly to the isotropic case Eq. (\ref{eq:RG2FinAppL}) this equation has a stable fixed point at $c_{x}^{*}c_{y}^{*} < c_{loc}^2$. In the infinite-size limit the steady state solution for conductance at that point can be approximated by 
 \begin{eqnarray}
\begin{pmatrix}
C_{x} \\
C_{y}
\end{pmatrix}  
=\frac{2c_{loc}^2}{c_{loc}^2-c_{x}^{*}c_{y}^{*}}\begin{pmatrix}
c_{x}^{*} \\
c_{y}^{*}
\end{pmatrix}   
\label{eq:RG2FinAppLAnis}
\end{eqnarray}
Conductance approaches infinity in the infinite-size limit for $c_{x}^{*}c_{y}^{*} \geq c_{loc}^2$. Consequently, the transition to the superdiffusive regime is defined as 
 \begin{eqnarray}
c_{loc}^2=c_{x}^{*}c_{y}^{*}.
\label{eq:AnisTrans}
\end{eqnarray}

\begin{figure}[h!]
\includegraphics[width=8cm]{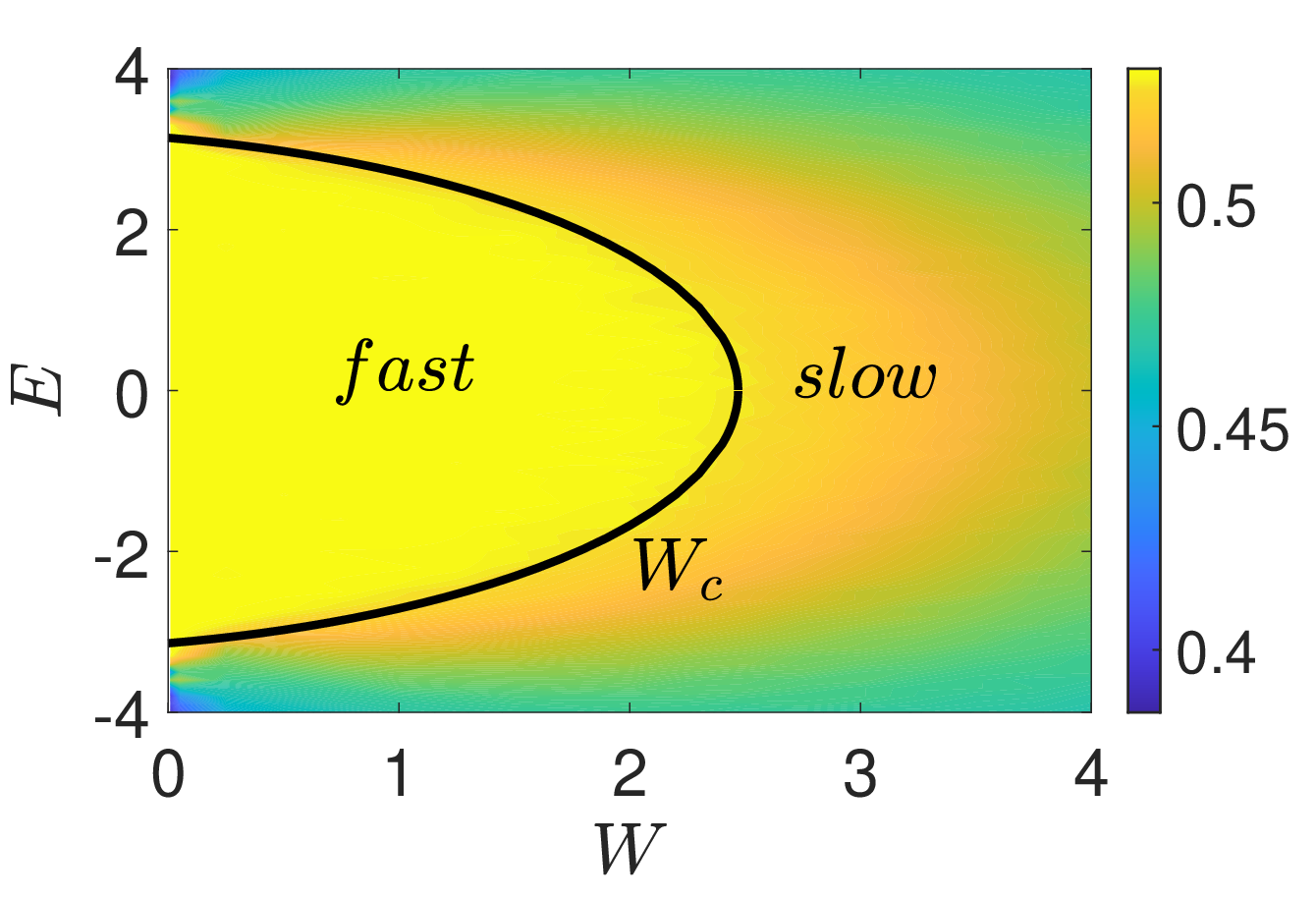}
\caption{Phase diagram of the 2D interacting dipoles model and the color code shows the average level-spacing ratio $\langle r \rangle$,  Eq. (\ref{eq:Rat})  characterizing the level statistics \cite{OganesyanHuse07}, evaluated for the system size $L=201$ and averaged over $1000$ realizations separately for each value of $W$. } The solid line shows the analytical predictions for the phase boundary in the infinite system determined using Eq. (\ref{eq:AnisTrans}).
\label{fig:LorPhD1}
\end{figure}

According to Eq. (\ref{eq:RG2FinAppL2}) conductance diverges logarithmically for $L \rightarrow \infty$  under the condition $c_{*} >  c_{loc}$ (the {\it fast} phase), while it remains finite otherwise (the {\it slow} phase). Setting $c_{*}=c_{loc}$ one can find critical disorder   separating phases.

For the dipole-dipole interaction the isotropic regime is realized only for the band center $E=0$, where the critical disorder is given by  $W_{c}=\pi/\sqrt{(1+\sqrt{5})/2}\approx 2.47$. With decreasing disorder the {\it fast} phase emerges first at that energy. For the anisotropic regime realized at $E \neq 0$ 
the transition emerges  at $c^{*}_{x}c^{*}_{y}=c_{loc}^2$, Eq. (\ref{eq:AnisTrans}). Using the analytical results, Eq. (\ref{eq:DrudeDipDip1}), for the logarithmic growing rates  $c^{*}_{x}$ and $c^{*}_{y}$ we determine the phase diagram depicted in Fig. \ref{fig:LorPhD1},  where solid lines indicate analytical predictions for the boundary between {\it slow}  and {\it fast} phases determined using Eq. (\ref{eq:AnisTrans}).

The dipole-dipole hopping  distance dependence  $r^{-2}$ emerges in high-dielectric-constant films. For the film  of the thickness $h$ possessing    a dielectric constant $\epsilon$ substantially exceeding that of the environment ($\epsilon_{env}$), $1/r^2$ hopping amplitude distance dependence  emerges for distances $r$ belonging to the domain $h<r<L_{max}=h\epsilon/\epsilon_{env}$ \cite{Smithe67Electricity,ab09Humidity} If the system is in the {\it fast} phase the conductance will grow within this domain reaching its maximum $C(L_{max}) \approx (c_{*}-c_{loc})\ln(\epsilon/\epsilon_{env})$. At longer distances  $L>L_{max}$ there is no long-range hopping contribution so the renormalization group equation takes the form $dC/d\ln(L)=-c_{loc}$ as for the short-range hopping. 
Then at distances exceeding $L_{max}$ a weak localization behavior of conductance is expected $C(L)=C(L_{max})-c_{loc}\ln(L/L_{max})$ until reaching the length  $l \approx h(\epsilon/\epsilon_{env})^{\frac{c_{*}}{c_{loc}}}$ where  $C(L) =0$.  At that scale  two dimensional Anderson localization is expected, so the size $l$ determines the localization length.  It seems to be an exciting experimental challenge to investigate the excitation displacements vs. time in high-dielectric-constant films to observe all three regimes.  For the hopping associated with elastic or RKKY interactions in isolated films there is no constraints like that for the dipole-dipole interaction so the superdiffusive behavior can be seen there at longer distances. 

\section{Numerical results}
\label{sec:Num}

Fast and slow phases should be distinguishable numerically and below we investigate the transition between them using exact diagonalization and considering the level statistics  (Sec. \ref{subsec:LStat}), fractal dimension (Sec. \ref{subsec:FractDim}) and transport kinetics (Sec. \ref{subsec:Transp}). Since the Hamiltonian matrix is not sparse, i. e. most of its elements are different from zero, the recently developed advanced diagonalization methods of large matrices \cite{Sierant2020LargeMatr} are not applicable to our problem of interest and the maximum size of the system is limited to $300\times 300$. Yet our results are quite consistent with the expectations of the analytical renormalization group theory. 

\begin{figure}[h!]
\centering
\includegraphics[width=8cm]{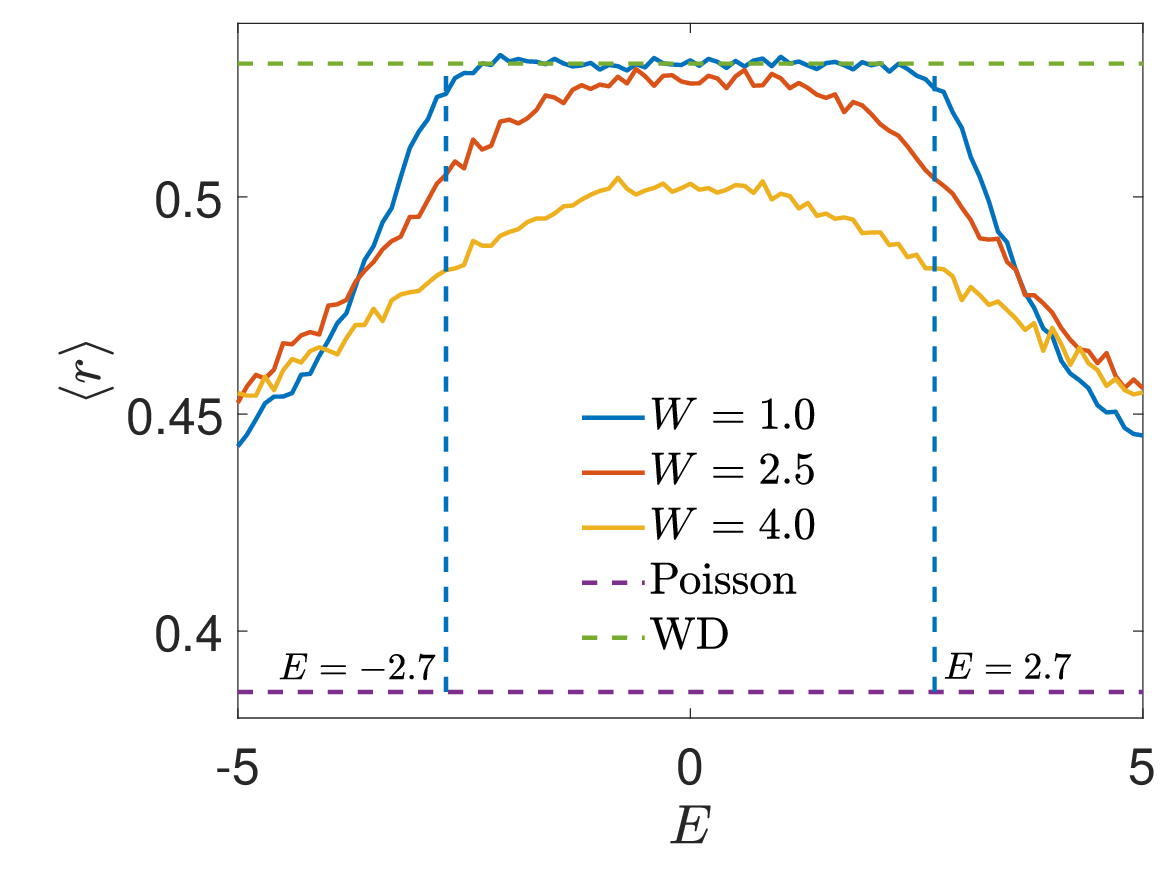}
\caption{Level-spacing-ratio $r$-statistics vs. energy for different disorder strengths $W$. The vertical dashed lines at  $E =\pm 2.7$ ($V_{0}=1$) indicate the fast phase borders  for the smallest disorder strength $W=1$.}
\label{fig:LStDip}
\end{figure}

\subsection{Level Statistics}
\label{subsec:LStat}

Energy level statistics is different for localized and delocalized states  \cite{ShklovskiiShapiro93}. For delocalized states it approaches the Wigner-Dyson random matrix energy level statistics due to energy level repulsion, while energies of localized states are independent of each other and can be characterized by a Poisson statistics.  Usually, Wigner-Dyson level statistics indicates ergodic behavior \cite{OganesyanHuse07,Atas2013LStatRandMatrRat}, see, however,  Ref.  \footnote{Though there are cases when Wigner-Dyson~\cite{Kravtsov_2015} and Poisson~\cite{Tang2022nonergodic,PhysRevLett.131.166401} level-spacing-ratio statistics coexists with the fractal extended eigenstates.}.


Eigenstates in the {\it slow} phase, characterized by a finite conductance,  are expected to have a fractal dimension $D$ reduced compared to a system dimension $d=2$ 
\cite{Levitov90,Mirlin19961DLR}. Consequently, the level repulsion should be reduced and we do not expect Wigner -  Dyson statistics and, consequently, ergodic behavior  in the {\it slow} phase. However,  one can expect it to appear in the {\it fast} phase similarly to the counterpart transition in $3D$ \cite{DengShlyap20163DDip}. This expectation turns out to be consistent with the numerical studies reported below. 

The level statistics is represented  in terms of the average ratio of the minimum to maximum of adjacent energy level splittings defined as \cite{OganesyanHuse07}
\begin{eqnarray}
\langle r \rangle =\left\langle \frac{{\rm min}(\delta_{n}, \delta_{n+1})}{{\rm max}(\delta_{n}, \delta_{n+1})}\right\rangle , ~ \delta_{n}=E_{n+1}-E_{n},
\label{eq:Rat}
\end{eqnarray}
where $E_{n}$ represent energies of eigenstates arranged in the ascending order. In the localized phase one has $\langle r \rangle = 2 \ln(2) - 1 \approx 0.386$, while in the delocalized, ergodic  phase characterized by  the Wigner-Dyson statistics $\langle r\rangle \approx 0.531$ \cite{Atas2013LStatRandMatrRat}.  

 In Fig. \ref{fig:LStDip}  we show the level statistics for the system of the size $L=291$ with the anisotropic dipole-dipole hopping Eq. (\ref{eq:dipdip})  and different disorder strengths averaged over $200$ realizations.with the energy resolution $\delta E = 0.1$. For the minimum disorder strength   $W=1$,  a substantial fraction of states with energies $|E|<2.7V_{0}$ indicated by the dashed line should belong to the fast phase (in all graphs we set $V_{0}=1$).  The average level spacing ratio parameter $\langle r \rangle$ approaches the Wigner-Dyson limit $0.53$ in this domain as we expected for the {\it fast} phase.  The intermediate disorder strength $W=2.5$ approximately corresponds to the last moment when the {\it fast} phase is present at $E=0$. For the  strongest disorder $W=4$  all states suppose to belong to the slow phase. It is visually clear in Fig. \ref{fig:LStDip}  that our numerical findings are consistent with the assumption of ergodic behavior in the {\it fast} phase and its lack in the {\it slow} phase. 
The data for the level statistics are also  presented in the phase  diagram Fig. \ref{fig:LorPhD1}. They  are consistent with the analytical results shown by the solid line.   The results are shown for the maximum probed system size $L=251$. The results for smaller sizes are quite similar to those in Fig. \ref{fig:LorPhD1} so we do not see any remarkable scaling of the level spacing ratio similarly to the earlier work \cite{ab21Scipost}. This is in contrast with the fractal dimension scaling reported below.

\subsection{ Fractal dimension}
\label{subsec:FractDim}

The numerical investigation of a fractal dimension reported in this section is targeted to verify a finite-size scaling of a  conductance. predicted by the renormalization group theory  Eq. (\ref{eq:RG2FinAppL2}) taking the  advantage of the earlier predicted connection of conductance and dimension  \cite{Falko95PrbFractDim2d}.  We define the fractal dimension using the informational dimension $D_{1}$  \cite{Renyi1959,nakayama2013fractal} that can be expressed in terms of the average eigenstate wavefunction Shannon entropy 
$\xi(L)=-\langle \sum_{i}|c_{i}|^2\ln(|c_{i}|^2)\rangle $, where the amplitudes $c_{i}$ represent eigenstate coefficients of the problem in the real-coordinate basis of sites $i$ with the energy $E$ close to zero ($-0.1<E<0.1$) and averaging is performed over all such states and different realizations of random potentials.

The informational dimension is connected to  the fractal dimension $D_{q}$ of the multifractal eigenstate defined for the specific exponent $q$ as \cite{ab21Scipost,cugliandolo2024multifractal}
\begin{eqnarray}
\langle \sum_{i}|c_{i}|^{2q}\rangle \propto L^{(1-q)D_{q}}. 
\label{eq:Multifractal}
\end{eqnarray}
It corresponds to the limit of $q\rightarrow 1$, where  the geometric averaging emerges naturally after both sides expansion in $1-q$ \cite{cugliandolo2024multifractal}.  In this limit the fractal dimension is less sensitive to fluctuations thus  reflecting a typical wavefunction behavior, which is our target.    

For numerical calculations we define a size-dependent informational dimension as    $D_{1}=d\xi/d\ln(L)$, cf. Ref.  \cite{MirlinEvers2000PowLawRandMatr}.  
The  dimension is estimated calculating the functions $\xi$ for the sequence of lengths $L_{1}$, $L_{2}$, ... $L_{n}$ arranged in ascending order and then numerically differentiating them.  This  yields $n-1$ estimates for fractal dimensions $D_{1}(l_{k})=(\xi(L_{k+1})-\xi(L_{k}))/\ln(L_{k+1}/L_{k})$ assigned to geometrically average sizes $l_{k}=\sqrt{L_{k}L_{k+1}}$ for  $k=1, 2...n-1$.  

Numerical results should be compared with analytical  estimates for fractal dimensions obtained using the generalized  theory Eq. (\ref{eq:RG2FinAppL2}) and the connection between the dimension and the system conductance. established within  the non-linear sigma model in Refs. \cite{Wegner1980,Altshuller1986Dim,
Falko95PrbFractDim2d,EversMirlin08}.  It was shown there  that the  informational dimension $D_{1}$ of a two-dimensional  system with a finite conductance is smaller than the system dimension $2$. The difference of dimensions is inversely proportional to  the dimensionless conductance at large conductance $C \gg c_{loc}$.    Consequently,  the fractal (informational) dimension can be expressed as  $D_{1}=2-\eta_{d} c_{loc}/C$ at $C \gg c_{loc}$.  Theory suggests \cite{Falko95PrbFractDim2d,EversMirlin08} $\eta_{d}=1$ for the non-linear sigma model  with a short-range hopping. 

\begin{figure}[h!]
\includegraphics[width=8cm]{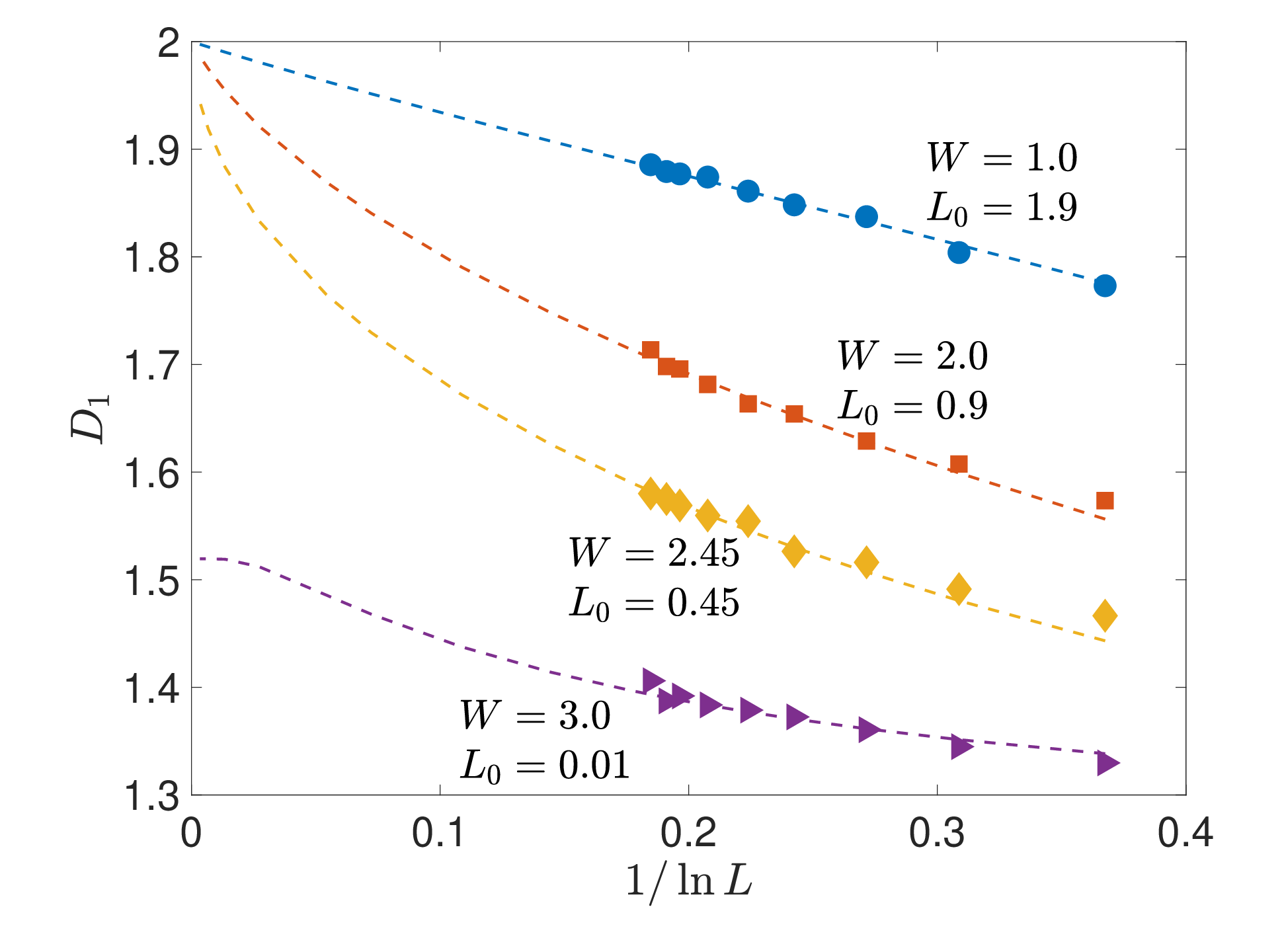}
\caption{Finite-size scaling of the fractal dimension vs. the predictions of the modified scaling theory of localization (see the text). The number of realizations is $40000$ for the minimum size $L=11$ decreasing with increasing the size to $2000$ for the maximum size $L=251$.}
\label{fig:FractDM}
\end{figure}

We were unable to fit the numerical data using  the analytical expression with $\eta_{d}=1$ (see Appendix \ref{sec:FrChoice}), 
 but obtained an excellent agreement between  analytical and numerical results   setting  $\eta_{d}=1.3$ ($D_{1}=2-1.3c_{loc}/C$).
 In Fig. \ref{fig:FractDM}  we present analytical results for the fractal dimension (solid line) together with its numerical estimate for the zero energy states of the system with the hopping due to the anisotropic  dipole-dipole interaction. 
The conductance was evaluated integrating  Eq. (\ref{eq:RG2FinAppL2}) and solving numerically the resulting transcendental equation that expresses size-dependent conductance as 
\begin{eqnarray}
C+C_{\infty}\ln\left(1-\frac{C}{C_{\infty}}\right)=(c_{*}-c_{loc})\ln\left(\frac{L}{L_{0}}\right). 
\label{eq:RGIntegr}
\end{eqnarray}
Here  $C_{\infty}=c_{*}c_{loc}/(c_{loc}-c_{*})$ represents the infinite-size limit of conductance in the {\it slow} phase  $c_{*} < c_{loc}$ 
 and $L_{0}$ is the unknown integration constant. We define this constant for each line shown in Fig. \ref{fig:FractDM}  minimizing the deviation of  analytical ($2-1.3c_{loc}/C$) and numerical estimates of fractal dimensions. at largest sizes where the theory is most relevant.    

The failure of the expression for $D_{1}$ with $\eta_{d}=1$ 
in the present model can be due to the long-range character of hopping.



 \subsection{Transport}
 \label{subsec:Transp}
 
Finally, we consider the particle transport that is  the main distinction of two phases.   Our goal here is just  to demonstrate that the superdiffusive transport, indeed, exists at small disorder strength  in a reasonable agreement  with the predictions of the analytical theory.  We do not attempt to compare the predictions of analytical theory with numerical results in detail because the transport includes many eigenstates with different energies and different transport rates that makes an accurate analytical consideration overcomplicated. 

As we explained earlier it is expected to be superdiffusive in the {\it fast} phase and diffusive in the {\it slow} phase.   For the initially localized particle its typical displacement $R$ increases with the time following a sort of diffusion law $R^2\propto C(R)t$ since conductance and diffusion coefficient are synonyms in two dimensions.  In  the fast phase  a conductance  $C(R)$  increases logarithmically with the size $R$ leading to a  superdiffusive behavior  in contrast to  the linear dependence  expected in the {\it slow} phase where conductance remains finite.  This expectation is consistent with  Ref.  \cite{Levitov90},  where the behavior $R \propto t^{1/d}$ was predicted for the {\it slow} phase in a $d$- dimensional system. For    
$d=2$ this is equivalent to the diffusive behavior in contrast to the subdiffusive behavior for $d=3$ \cite{Levitov90} or the superdiffusive behavior for $d=1$ \cite{Mirlin19961DLR}. 

\begin{figure}[h!]
\includegraphics[width=8cm]{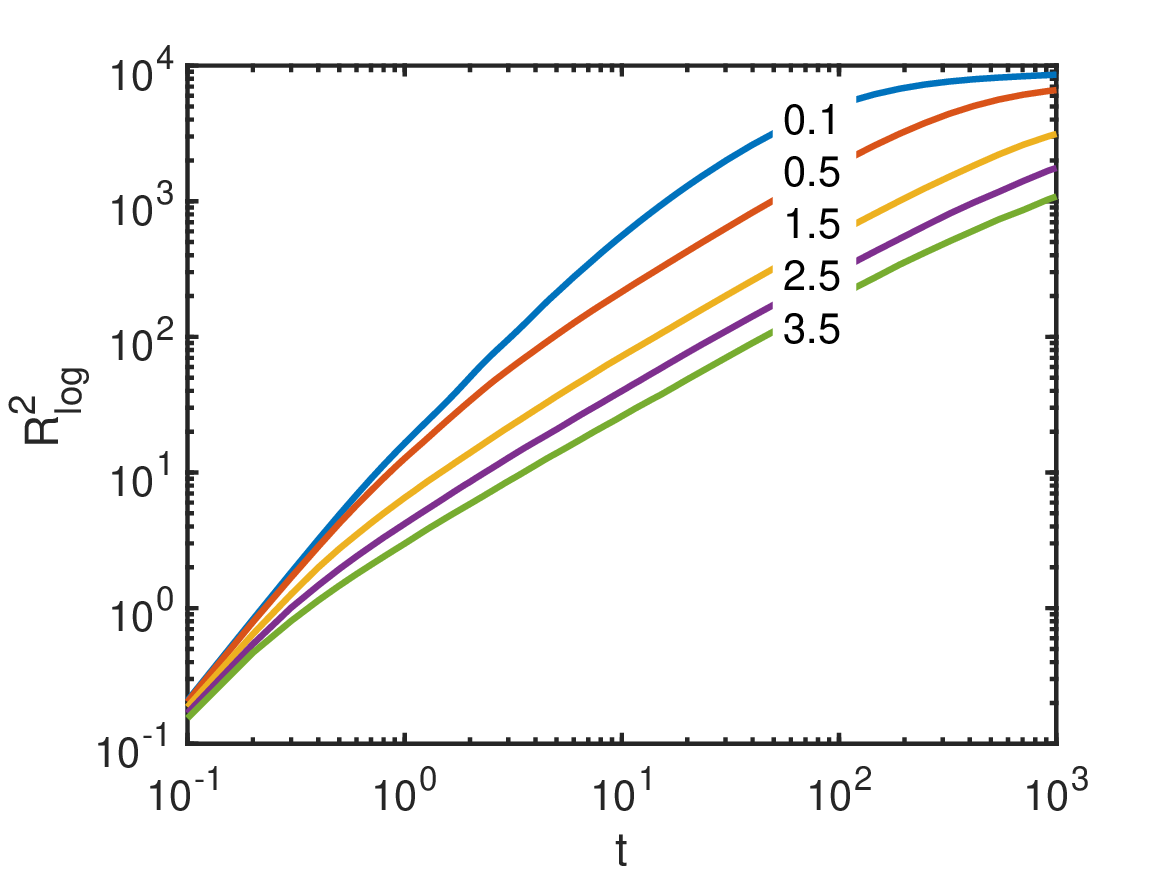}
\caption{Time dependence of typical squared displacement at different disorder strengths $W$ indicated  near each graph for the system size $L=251$, averaged over $300$ realizations.  }
\label{fig:DispTDep}
\end{figure}


To verify these expectations we  investigate the time evolution of the  state $c_{i}$ initially ($t=0$) localized in the origin $i=0$ ($c_{i}(0)=\delta_{i0}$, where $\delta_{i0}$ is the Kronecker symbol).  We set a random potential in the origin to zero ($\phi_{0}=0$, see Eq. (\ref{eq:HamGen})) to have the average energy of the state of interest equal to zero $0$, where delocalization emerges in the maximum extent. Different strengths of random potentials were investigated  including $W=0.1$, $0.5$  and $1.5$ for the {\it fast} phase, $W=2.5$ for the transition point  and $W=3.5$ for the {\it slow}  phase. For $W=0.1$, $0.5$ or $1.5$ the fast phase is realized for the majority of the states, except for a small fraction of the ``slow" states in the tails of the spectrum that don't affect the particle transport. The choice of a zero random potential in the  origin where the particle was placed at $t=0$ reduces the contribution of  {\it slow} tail states to the wavefunction evolution.  

\begin{figure}[h!]
\includegraphics[width=8cm]{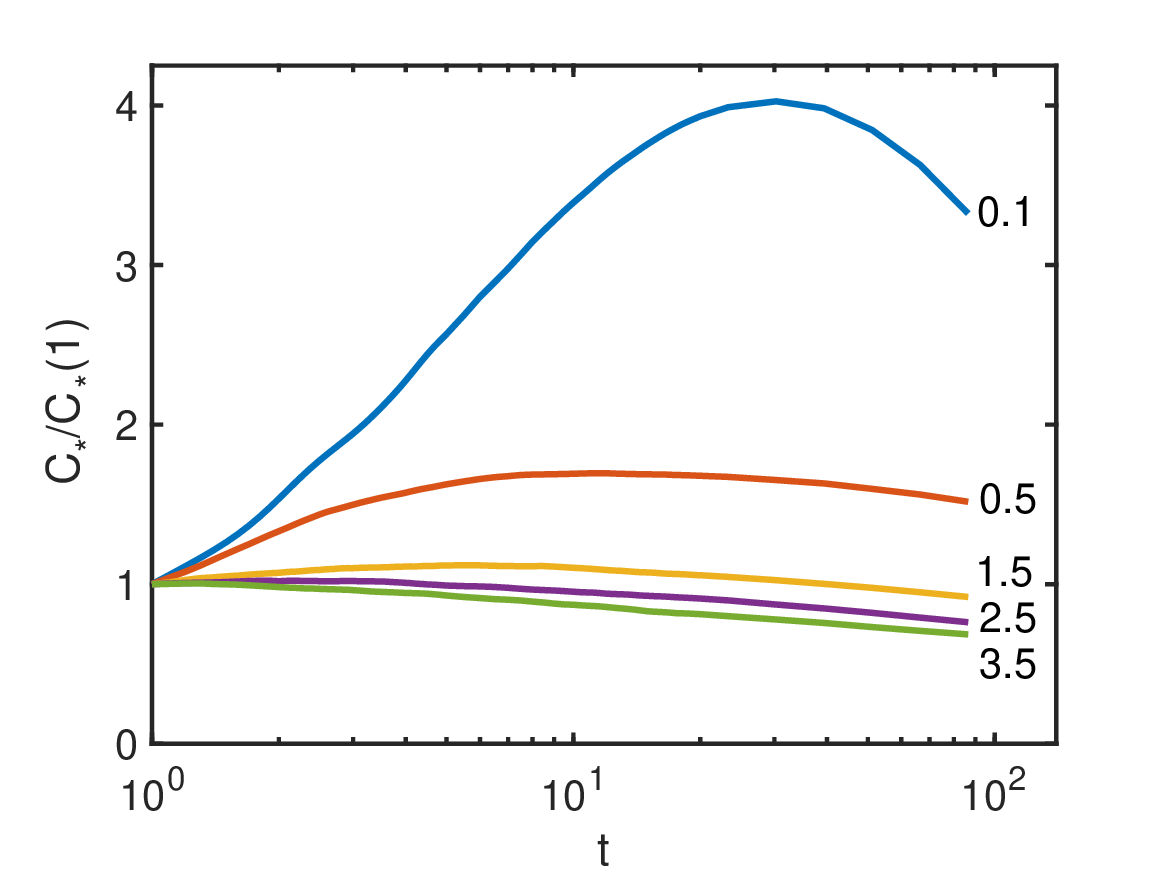}
\caption{Time dependence of relative conductance $C_{*}/C_{*}(1)$  ($C_{*}(t)=R_{\rm log}^2(t)/t$) at different disorder strengths $W$ indicated on the right of each graph for the same conditions as in Fig.  \ref{fig:DispTDep}.  }
\label{fig:CondTDep}
\end{figure}


The particle transport has been characterized using the logarithmically average  displacement $R_{\rm log}(t)$ (excluding the origin), defined as 
\begin{eqnarray}
\ln\left(R_{\rm log}(t)^2\right) = \sum_{i\neq 0} |c_{i}(t)|^2 \ln(R_{i}^2). 
\label{eq:rsq}
\end{eqnarray}
We do not consider the most often used mean squared  displacement  because the power-law tails of the wavefunction can lead to the overestimate  of the actual move.   Indeed, even at short times a mean squared displacement diverges with the system size because of the wavefunction asymptotic behavior $\psi(R) \propto R^{-2}$ emerging in the first order perturbation theory in a hopping. Although at short times  the particle is localized  nearby the origin $R=0$,  the average squared displacement $\int d^{2}\mathbf{R}|\psi(R)|^2R^2$   diverges logarithmically for $\psi(R) \propto R^{-2}$ for the system of an infinite size, while logarithmic averaging does not lead to any divergence. at a finite time.  

The results of the calculations for $R_{\rm log}(t)^2$ vs. $t$ are shown in Fig. \ref{fig:DispTDep}.  Based on these results it is still difficult to characterize the transport because at very short times $t < 1$, $1/W$ one has $|c_{i}(t)| \approx V_{i}t$ ($V_{i}$ is the coupling strength of the initial site and the site $i$) leading to $R_{\rm log}(t) \propto t$ , while the saturation at around the system size takes place at long times $t>100$.  To  focus on the  relevant time domain for the superdiffusive transport we restrict our consideration to times $t_{min}<t<t_{max}$ with   $t_{min}=1$ and  $t_{max}=100$.  The time dependence of the relative conductance $C_{*}(t)/C_{*}(1)$ ($C_{*}(t)=R_{\rm log}(t)^2/t$ in that time domain in shown in Fig.  \ref{fig:CondTDep} for various disorder strengths.  Relative conductance is used since we are interested in the time dependence of the diffusion coefficient (conductance) rather than its absolute value. 
If the transport is superdiffusive this effective conductance should increase with the time. logarithmically, saturating at long times due to a finite-size effects.  

 According to Fig.   \ref{fig:CondTDep}  the superdiffusive transport, indeed, emerges for the weakest  disorder at $W \leq 1.5$ in accord with the theoretical expectations.  At short times the conductance time dependence is consistent with the expected logarithmic growing  for $W=0.1$ and $W=0.5$, while for $W=1.5$ the growing domain is too narrow to make any conclusions possibly due to finite-size effects.  For stronger disorder the conductance slowly decreases with the time.   
 At the transition point $W \approx 2.5$ corresponding to the zero energy, Eq. (\ref{eq:RG2FinAppL2}) predicts the superdiffusive behavior $C \propto \sqrt{\ln(L)}$, that we do not see, possibly because of  dominating contributions of slow states with energies different from $0$.  The reduction of the diffusion coefficient (conductance) with the time (size) for $W>1.5$ can be due to the renormalization of coupling constant  occurring for the dipolar hopping and dipoles oriented within the same direction for  strong disorder \cite{ab96DipMomRen}.  Although disorder is not  strong, some reduction of the effective coupling constant can still take place.




\section{Conclusions}
\label{sec:Concl}

We show the emergence of a superdiffusive {\it fast} phase in two-dimensional Anderson model with long-range hopping $V(r) \propto r^{-2}$, possessing the time-reversal symmetry, at  sufficiently small disorder. The {\it fast} phase is characterized by delocalized, ergodic  eigenstates occupying the whole space and fostering the superdiffusive transport. The complementary {\it slow} phase is non-ergodic. In this phase eigenstates are delocalized, while their  fractal dimension is less than $2$. The transport there is expected to be diffusive but  restricted to the maximum displacement substantially smaller than the system size. 

The conductance of the system is finite in the {\it slow phase} and infinite in the {\it fast} phase. It continuously approaches infinity in the transition point in contrast to the other known  localization-delocalization transitions  \cite{LeeRamakrishnan1985RevModPhys,Aleiner112D}. The boundary between two phases is determined analytically, which is unprecedented for Anderson localization problem with the only exception of the celebrated self-consistent theory of localization valid for the Bethe lattice \cite{AbouChacra73}. 

\begin{acknowledgments} 
{\it Acknowledgement.} A. B. acknowledges the support by Carrol Lavin Bernick Foundation Research Grant (2020-2021),  NSF CHE-2201027 grant and LINK Program of the NSF EPSCOR (award number OIA-1946231) and Louisiana Board of Regents. 
  X.D acknowledges the support by the Federal Ministry of Education and Research of Germany (BMBF) in the framework of DAQC. I. M. K. acknowledges the European Research Council under the European Union’s Seventh Framework Program Synergy HERO SYG-18 810451.
\end{acknowledgments}

\bibliography{MBL}

\appendix



\section{Fourier transforms of hopping amplitudes} 
\label{sec:Modell}

The anisotropic dipole-dipole interaction, responsible for the hopping in the model under consideration, is 
defined by Eq. (\ref{eq:dipdip}).  Here we calculate numerically  its  Fourier transform   $V(\mathbf{q})=\sum_{j}V_{ij}e^{i\mathbf{qr}_{ij}}$  needed to  evaluate the classical  conductance, Eq. (\ref{eq:DrudeDipDip1}).

\begin{figure}[h]
\includegraphics[scale=0.4]{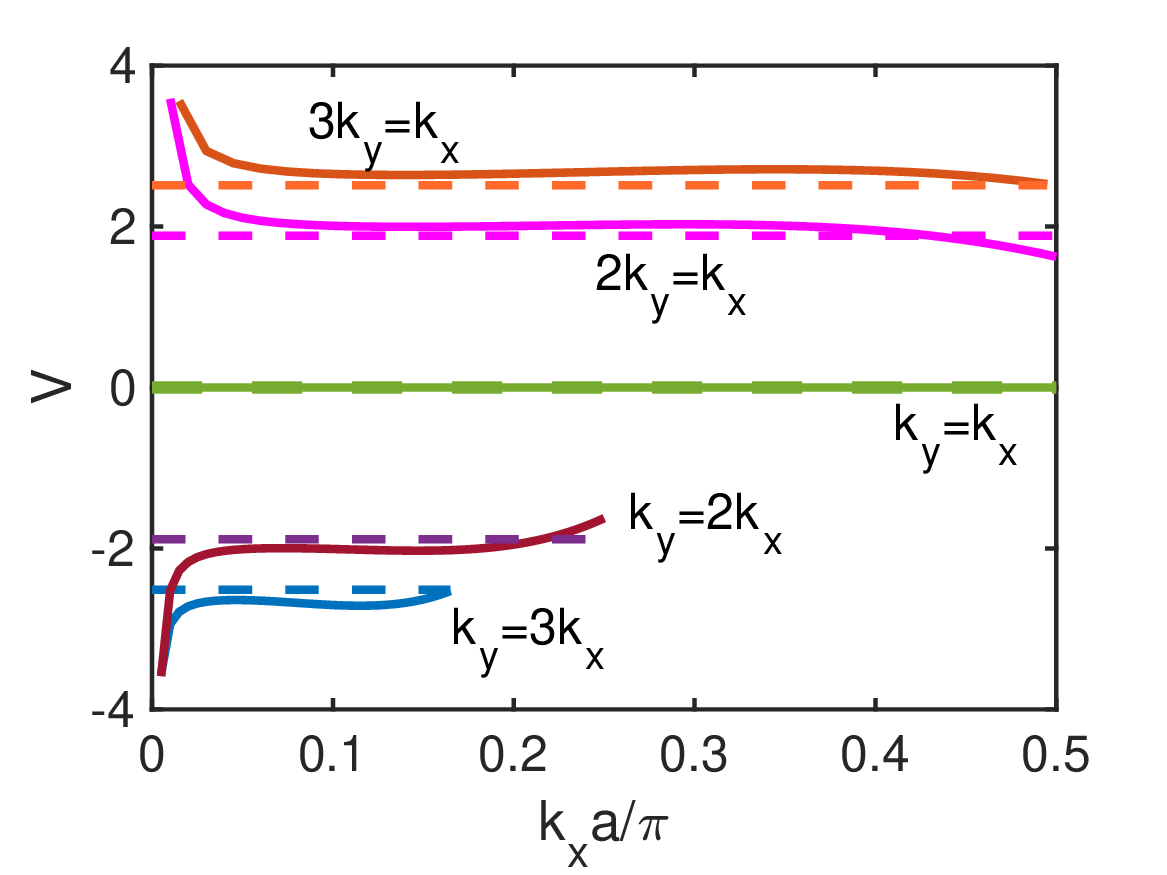}
\caption{\small   Comparison of analytical and numerical Fourier transforms for dipole-dipole interaction (at fixed ratios $k_{y}/k_{x}$). Numerical Fourier transform is evaluated for the system of the size $L=100$. Analytical results is shown by dashed lines and numerical results are shown by solid lines.  } 
\label{fig:Four}
\end{figure}

It turns out that dipole-dipole  interaction Fourier transform can be well represented by its continuous limit given by 
\begin{eqnarray}
V(\mathbf{q})=2V_{0}\pi \frac{q_{y}^2-q_{x}^2}{q^2}, 
\label{eq:DipFour}
\end{eqnarray}
This approximation works reasonably well for the periodic square lattice of the size  $L=100$ as illustrated in Fig. \ref{fig:Four}. It becomes exact for the system size approaching infinity since the logarithmic divergence of the  classical conductance emerges at  $q\rightarrow 0$ corresponding to long distances. In this limit the regular correction to Eq.  (\ref{eq:DipFour}) disappears because the sum of all dipole-dipole interactions is zero.

\section{Connection of conductance and informational  dimension.}
\label{sec:FrChoice}

\begin{figure}[h]
\includegraphics[scale=0.25]{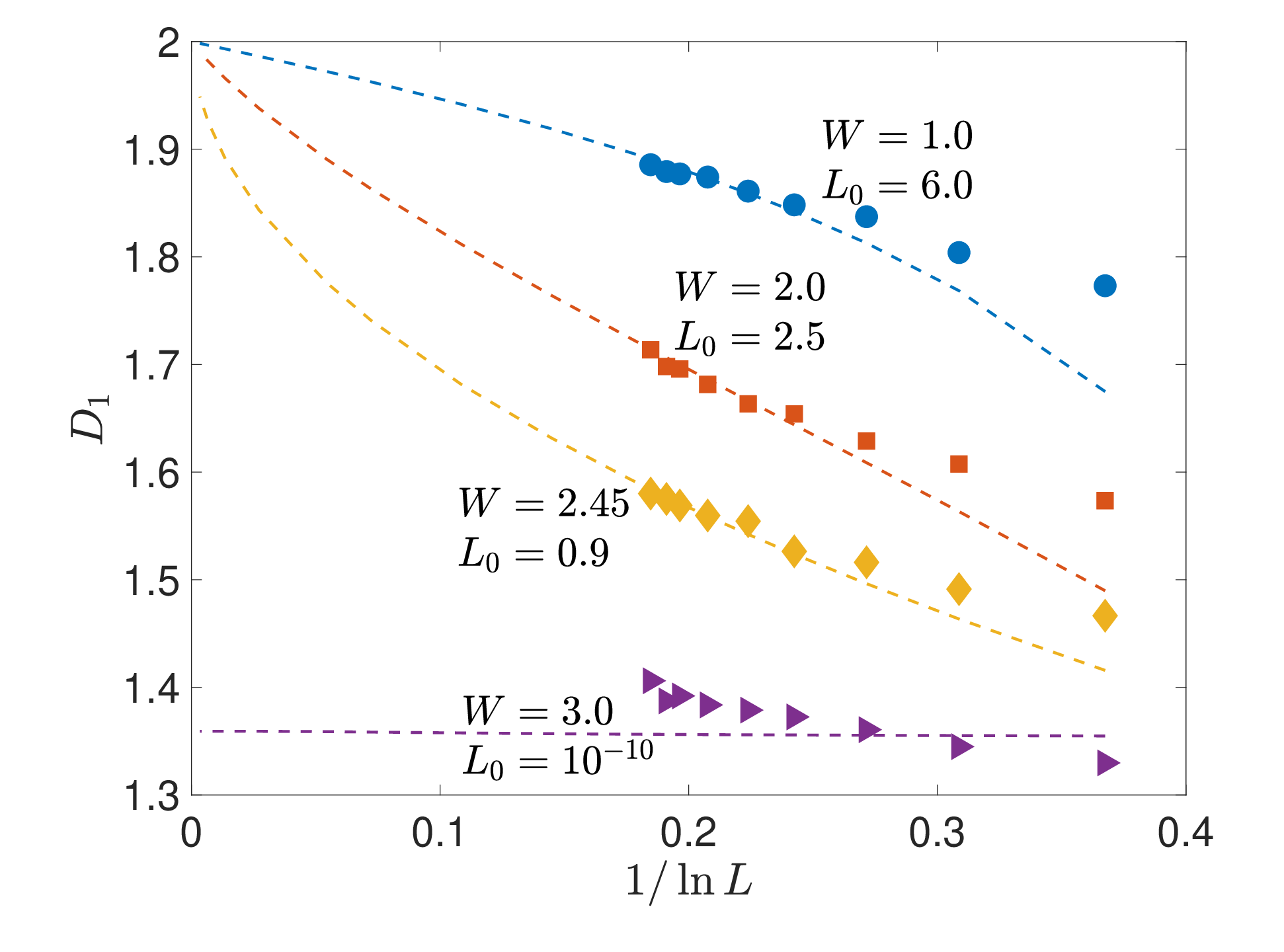}
\caption{\small   Comparison of the numerical estimate of eigenstate fractal dimensions with the analytical theory of Ref. \cite{Aleiner112D}.   
The only exception is the case of $W=3$ for the present graph, where we cannot fit those data even choosing the minimum possible value of  $L_{0} \rightarrow 0$, corresponding to the infinite-length limit of the fractal dimension.,  as shown by the dashed line..} 
\label{fig:FailTha}
\end{figure} 

Here we show that the numerically calculated fractal dimension $D_{1}$ cannot be fitted by the analytical theory \cite{Falko95PrbFractDim2d}  predicting the dependence of this dimension on the conductance in the form $D_{1}=2-\frac{c_{loc}}{C}$ in contrast with the dependence $D_{1}=2-1.3\frac{c_{loc}}{C}$ used in the main text. We evaluated a conductance using the scaling equations derived in the present work and in Ref.  \cite{Aleiner112D}, which can be both written in the form 
 \begin{eqnarray}
\frac{d\ln(C)}{d\ln(L)}=\frac{c_{*}-c_{loc}}{C} +\zeta \frac{c_{*}c_{loc}}{C^2},  
\label{eq:ScThMainApp}
\end{eqnarray}
with  $\zeta=1$ for the present work and  $\zeta=1/2$ for Ref. \cite{Aleiner112D}.

The conductance $C$ is found integrating Eq.  (\ref{eq:ScThMainApp}) as (see Eq. (\ref{eq:RGIntegr}) in the main text) 
\begin{eqnarray}
C+C_{\infty}\ln\left(1-\frac{C}{C_{\infty}}\right)=(c_{*}-c_{loc})\ln\left(\frac{L}{L_{0}}\right), 
\label{eq:RGIntegrEta}
\\
C_{\infty}=\zeta \frac{c_{*}c_{loc}}{c_{loc}-c_{*}}.
\nonumber
\end{eqnarray}
We choose  the optimum parameter $L_{0}$ in Eq. (\ref{eq:RGIntegrEta}) as in the main text  minimizing the deviation of analytical and numerical results  at largest sizes.  


\begin{figure}[h]
\includegraphics[scale=0.25]{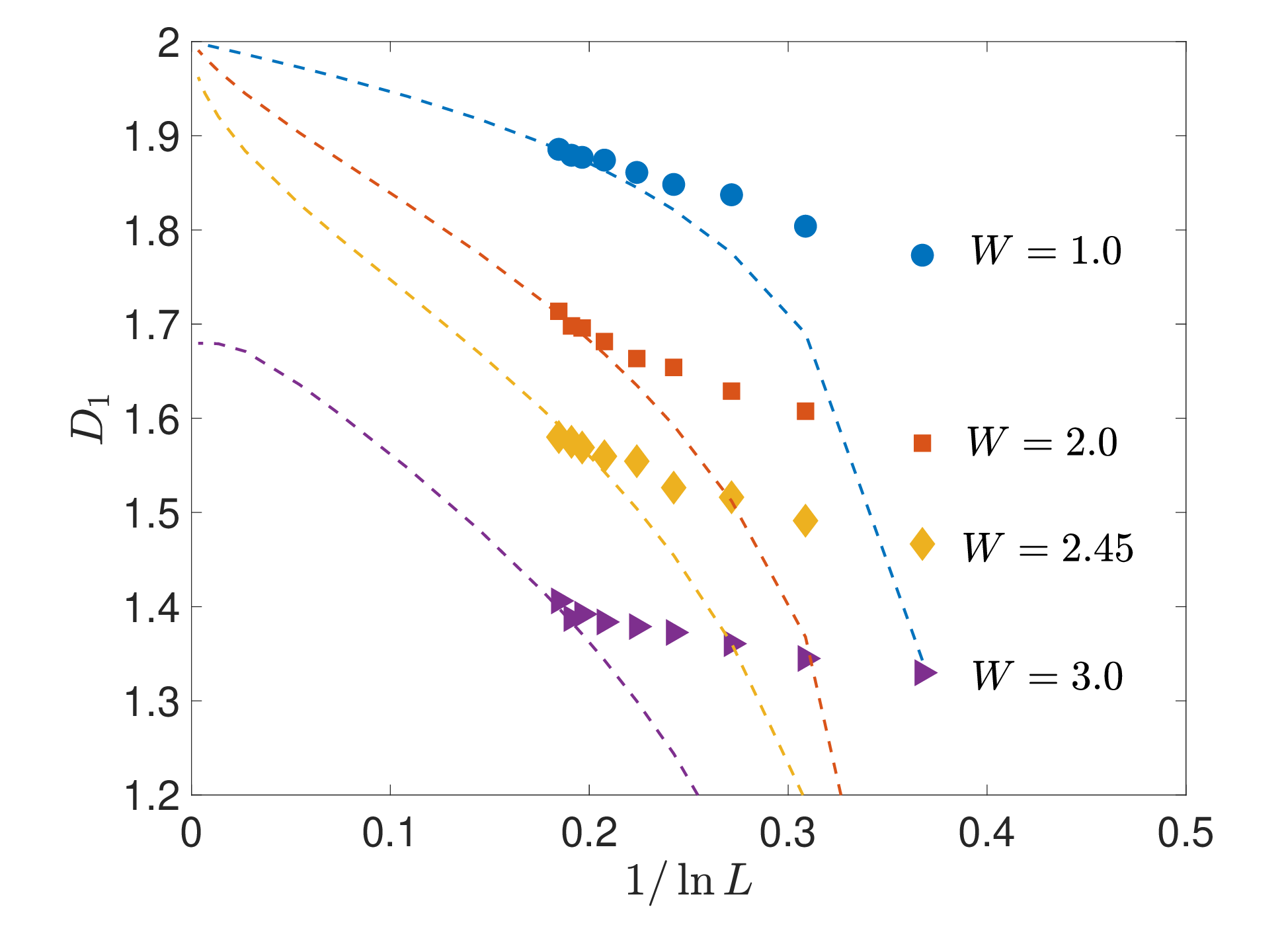}
\caption{\small   \small   Comparison of the numerical estimate of eigenstate  fractal dimensions with the analytical theory of the present work.  We always used $L_{0}=10$. } 
\label{fig:FailThb}
\end{figure} 

In Figs. \ref{fig:FailTha}, \ref{fig:FailThb} we show the comparison of the numerical results for the dimension $D_{1}$  (the numbers of realizations are as in Fig. \ref{fig:FractDM}) and the analytical theory $2-c_{loc}/C$ for two choices of the parameter $\zeta=1/2$ and $1$, respectively.  
It is clear from Figs. \ref{fig:FailTha}, \ref{fig:FailThb}   that in both cases the analytical theory does not provide an  acceptable fit of the data. However,  if we set $D_{1}=2-1.3c_{loc}/C$, and employ Eq. (\ref{eq:ScThMainApp}) with the parameter $\zeta=1$,  then we get an almost perfect agreement of numerical and analytical results  as reported in the main text,  Fig. \ref{fig:FractDM}.  

\end{document}